# Superior lattice thermal conductance of single layer borophene

Hangbo Zhou[#], Yongqing Cai[#], Gang Zhang[1], and Yong-Wei Zhang

Institute of High Performance Computing, A*STAR, Singapore 138632

#Both authors contributed equally to the work. Email: zhangg@ihpc.a-star.edu.sg

**Abstract**

By way of the nonequilibrium Green's function simulations and first-principles calculations, we report that borophene, a single layer of boron atoms that was fabricated recently, possesses an extraordinarily high lattice thermal conductance in the ballistic transport regime, which even exceeds graphene. In addition to the obvious reasons of light mass and strong bonding of boron atoms, the superior thermal conductance is mainly rooted in its strong structural anisotropy and unusual phonon transmission. For low-frequency phonons, the phonon transmission within borophene is nearly isotropic, similar to that of graphene. For high-frequency phonons, however, the transmission is one-dimensional, that is, all the phonons travel in one direction, giving rise to its ultra-high thermal conductance. The present study suggests that borophene is promising for applications in efficient heat dissipation and thermal management, and also an ideal material for revealing fundamentals of dimensionality effect on phonon transport in ballistic regime.

## I. Introduction

Thermal transport is known to be one of the major forms of energy transfer in nature. The well-known Fourier's law, a pioneering work by Fourier in 1820s, is a milestone in the study of thermal conduction. Over the past few decades, due to the fast development of nanotechnology, and the rapid rise of novel concepts in manipulating and controlling heat energy transfer[1], the understanding on the phonon-derived thermal transport in low-dimensional systems has progressed at a remarkable pace. With regard to thermodynamics and thermal transport properties, carbon-based materials are unique in the sense that carbon atoms can form strong covalent bonds with sp, $sp^2$, and $sp^3$ hybridizations. Owing to the strong covalent bonding, carbon allotropes, including three-dimensional diamond, two-dimensional (2D) graphene, and quasi-one-dimensional carbon nanotubes, have attracted much interest due to their stability and various fascinating properties[2]. Although the electronic properties of these carbon allotropes are notably different, for instance, *$sp^3$* diamond is a wide band gap insulator while *$sp^2$* graphene is a semimetal with extraordinarily high electron mobility; all of them have one common characteristic, *i.e.*, the ultra-high thermal conductivity[3-9]. For instance, diamond possesses the highest known lattice thermal conductivity in three-dimensional materials[10, 11]. In terms of nanostructures, carbon nanotube has a thermal conductivity higher than 3000W/Km[12, 13], which is the highest among all the one-dimensional materials. In the family of 2D materials, graphene possesses the highest reported thermal conductivity at room temperature[14-18], which is 2~3 orders higher than that of other 2D materials, such as $MoS_2$[19-25] and phosphorene[26-31].

The underlying reasons for the highest thermal conductivity of carbon-based materials are mainly due to their light atomic mass, short bond length, and strong covalent bonds between carbon atoms. Then, an interesting question arises naturally: Is there any material that its thermal conduction is even better than its carbon counterpart in the same dimension? Possible candidates could be a material made from the element sitting before carbon in the periodic table, and yet possessing comparable or even stronger bond strength. However, hydrogen is in gas phase. Lithium and beryllium are in the form of metallic bonds at room temperature, whose bond strengths are much weaker than covalent bond and thus their lattice thermal conductance is expected to be lower. Therefore boron is the only possible choice since it has the potential to form strong covalent bonds. However, the three-dimensional structures of boron are either amorphous or polymorphous, thus significantly limiting its thermal conduction.

Recently, a 2D sheet of boron atoms, called borophene, has attracted a plenty of research interest[32-34]. Remarkably, monolayer borophene has been grown on metal surface[35, 36] and it has been shown that it is stable against oxidization[36]. Undoubtedly, the experimental realization of borophene raises the hope that it might serve as a high-speed pathway for phonon transport, due to its short covalent bonds and light atomic mass. Although the thermal conductivity[37, 38] and thermal conductance[39] of borophene have been explored, the ballistic phonon transport mechanisms and properties of borophene are still largely unknown, and thus a solid understanding is needed in order to provide guidance for the exploration of its applications.

In this work, by using first-principles calculations and non-equilibrium Green's function (NEGF) technique, we investigate the lattice thermal transport properties of borophene for both hexagonal model and β12 model. It is found that the thermal conductance of borophene is highly anisotropic. At room temperature, the lowest thermal conductance is 62% of the highest one in the hexagonal model, while it is only 30% in the β12 model. Remarkably, the highest thermal conductance at room temperature reaches 7.87nW/(K·nm$^2$) and 10.97 nW/(K·nm$^2$) for hexagonal and β12 structures, respectively, which are obviously higher than that of graphene.

## II. Computational Methods

**DFT calculation**

The interatomic force constants (IFC) were calculated within the frameworks of density functional perturbation theory (DFPT) by using the Quantum-Espresso package[40]. The generalized gradient approximation (GGA) with the exchange-correlation functional in the Perdew-Burke-Ernzerhof parametrization was adopted together with ultrasoft pseudopotential. For borophene calculation, the first Brillouin zone was sampled with a 48×28×1 and 13×11×1 Monkhorst-Pack (MP) meshes for electrons and phonons, respectively. Cutoff energies of 40 Ry for plane wave and 600 Ry for electronic density were adopted. For the calculation of graphene, we use MP meshes of 21×21×1 (electrons) and 8×8×1 (phonons). The atomic structures and lattice constants were fully relaxed until the forces on each atom became smaller than $1\times10^{-6}$ eV/Å. In the calculation of phonon dispersion relation, the first step is to check the force constant to valid the acoustic sum rule (ASR), which requires that the frequency of acoustic mode at $k=0$ should vanish. This rule is originated from translational invariance, which requires that the net force applied to each atom equals to zero for rigid translation of atoms associated with the acoustic phonons at the zone-center. In this work, the phonon dispersion relation is also calculated using Quantum-Espresso package[40].

**Non-equilibrium Green's function**

By using the DFPT calculation, we obtain the interatomic force constants and hence we can write the Hamiltonian of lattice vibration as a collection of harmonic oscillators as:

$$H = \frac{1}{2m}p^T p + \frac{1}{2}u^T K u \quad (1)$$

where $u$ is a column vector denoting the atomic displacements in all degrees of freedom, and $p$ is the column vector of momentum and conjugate of $u$. The atomic mass is $m = 10.8\text{u}$ for boron and $K$ is the force constant matrix obtained from DFPT. We consider that borophene is infinitely large hence only the thermal conductance per cross-sectional area is meaningful, which can be obtained from Landauer formula:

$$\sigma/S = k_B \int_0^\infty \frac{d\omega}{2\pi S} \hbar\omega \mathcal{T}[\omega] \frac{\partial f}{\partial T} \quad (2)$$

where $f[\omega] = 1/(e^{\hbar\omega/k_B T} - 1)$ is the Bose-Einstein distribution, $S$ is the cross-sectional area and $\mathcal{T}[\omega]$ is the total transmission coefficient. By making use of the periodicity in the transverse direction of the transport, we can obtain the transmission coefficients by integrating over the transverse modes:

$$\mathcal{T}[\omega]/S = \int_{BZ} \frac{d\mathbf{k}_\parallel}{2\pi\delta} \Xi[\omega, \mathbf{k}_\parallel] \quad (3)$$

where $\delta$ is the thickness of the materials, $\mathbf{k}_\parallel$ is the wavevector in the transverse direction of the transport, and $\Xi[\omega, \mathbf{k}_\parallel]$ is the transmission coefficeints of phonons with energy $\omega$ and transverse wavevector $\mathbf{k}_\parallel$, and can be evaluated according to the force constant matrix within the non-equilibrium Green's function framework as demonstrated in references[41, 42].

**III. Results and Discussions**

**Phonon dispersion relation of hexagonal borophene**

The lattice structure of hexagonal borophene shown in Figure 1(a) is the typical monolayer hexagonal boron sheet, which has been reported experimentally[35]. In this structure, two planes of rectangular lattice form an out-of-plane buckled structure, with the buckling height of 0.89 Å. The lattice constants are 1.612 Å in the x direction (the ridge direction) and 2.869 Å in the y direction (the across-ridge direction), respectively. By using DFPT, we obtained the IFC and hence the phonon dispersion relation of borophene, which is shown in Figure 1(b). With two atoms in the unit cell, monolayer borophene has three acoustic branches: the in-plane transverse acoustic (TA) mode, the longitudinal acoustic (LA) mode, and the out-of-plane acoustic (ZA) mode, and three optical phonon branches. As shown in Figure 1(b), the LA and TA modes around the zone center along the y direction show a lower group velocity than those along the x direction. In addition, the phonon bands of the optical modes along the y direction are generally more flat than that along the x direction. Such phenomenon implies that the bond strength of borophene is highly anisotropic. It is worth mentioning that there exist slightly imaginary modes near Γ point, suggesting that the monolayer borophene is unstable against long wavelength periodic distortions, which is consistent with previously calculated phonon dispersion[35]. Similar imaginary modes were also observed in graphene nanoribbon[43]. With supporting by a proper substrate, the instability of the structure may be suppressed. It should be noted, however, that despite this small numerical uncertainty in the thermal conductance at extremely low temperature, such slightly imaginary modes should not significantly affect the accuracy in the thermal conductance calculations[43].

From Figure 1(b), it is seen that there is a cut-off frequency of ~1350cm$^{-1}$ in the phonon spectrum of borophene, which is comparable to that of graphene (1600cm$^{-1}$)[6]. This indicates that borophene possesses a strong interatomic bonding. Similar to graphene, the strong covalent bonding in borophene suggests that it has a high Debye temperature[44]. Therefore, the temperature for the thermal excitation of the highest frequency phonon mode in borophene is comparable to that of graphene, which is about 2000K. Hence room temperature (300K) can be considered as low temperature with respect to its high Debye temperature, and therefore, lattice vibrations should be treated quantum-mechanically, in the same way as done in carbon nanotubes and graphene[43, 45-48].

**Ultra-high quantum thermal conductance in borophene**

Due to the high Debye temperature (2000K), the phonon population is low at room temperature, which suppresses the possibility of phonon-phonon inelastic scattering events. As a result, all lattice vibrations in borophene can be decomposed into normal modes, which are nearly independent of each other[49, 50]. Furthermore, when the sizes of a material decrease, in particular, when the length scale is comparable or smaller than the phonon mean free path, the inelastic scattering process will be further suppressed. Thus, the thermal transport is quasi-ballistic or ballistic. In this regime, there are many striking phenomena which cannot be explained by phonon diffusion theory, for example, the quantization of thermal conductance[51, 52], and the universal phonon-transmission fluctuation[53]. Moreover, a study of such a system necessarily requires the incorporation of quantum effects into thermal transport analysis. The non-equilibrium Green's function method is able to exactly solve the quantum ballistic transport problems, and therefore it is a powerful tool for studying low temperature transport properties when non-linear interactions are weak. In fact, it has been widely used to study the heat transport problems in carbon-based materials[43, 45-48].

Here, we employ the non-equilibrium Green's function (NEGF) formalism to investigate thermal conductance of borophene. Since a material with a larger width $W$ possesses more phonon transport channels and thus larger thermal conductance, the scaled thermal conductance, defined as the thermal conductance per unit area, is introduced to characterize the thermal transport properties. Herein, the cross-sectional area $S$ is defined to be $S = W \times \delta$. For borophene, the buckling height is 0.89Å. Hence, its thickness is chosen as 2.70Å, by adding the diameter of boron atom of 1.81 Å.

The thermal conductance per cross-sectional area as a function of temperature along both the x and y directions is shown in Figure 2. The increasing trend of thermal conductance with temperature is general for ballistic transport since more phonon branches are activated at a higher temperature[43, 45-48]. As temperature increases, the curve of thermal conductance gradually levels off. Clearly, the room temperature (300K) is far below the temperature of 900k, corresponding to the level-off thermal conductance of borophene, further verifying that at room temperature, the number of excited phonons is the major factor that dictates the thermal conductance.

From Figure 2, it is seen that borophene has a strong anisotropy in thermal conductance. At room temperature, the lattice thermal conductance in the x direction is 7.87nWK$^{-1}$ nm$^{-2}$ while that in the y direction is 4.89nWK$^{-1}$ nm$^{-2}$. Hence, the thermal conductance in the x direction is around two times of that in the y direction. Using the same method, we find that the lattice thermal conductance of graphene at room temperature is 5.09nWK$^{-1}$nm$^{-2}$ (the green line in Figure 2), which is consistent with previous results[48, 54]. Remarkably, the quantum thermal conductance in the x direction is 55% larger than that of graphene. Even in the y direction, the room temperature quantum thermal conductance of borophene is still comparable to that of graphene. We would like to emphasize that this surprising feature of extraordinarily high thermal conductance over graphene has not been observed in other 2D materials.

Conventionally, the atomic mass, interatomic forces and atomic density are believed to be the major factors affecting the lattice thermal conductance. However, these factors cannot explain the much higher thermal conductance of borophene. For instance, the boron atom is only slightly lighter than carbon atom, and the number of atoms per unit area of borophene in the x direction (13.7atoms/nm$^2$) is even lower than that of graphene (13.9atoms/nm$^2$). Moreover, the Young's module along the x direction of borophene (398Gpa nm) is not remarkably different from that of graphene (340Gpa nm)[35], signifying that the strength of B-B bond and C-C bond are comparable. Hence, new mechanisms that lead to the superior quantum thermal conductance of borophene should exist.

**Phonon transmission coefficients**

In order to understand the underlying mechanisms, we investigate the transmission coefficients of the phonons. The transmission coefficient is an important quantity in the NEGF formalism in describing the transmission probability of phonon modes. From the transmission coefficients, we can analyze the contribution of each mode to the lattice thermal conductance and identify the underlying origin why borophene possesses such an extraordinarily high thermal conductance that exceeds graphene.

The transmission coefficients of phonons at different frequencies are shown in Figure 3(a). It is seen that in the low frequency regime (below 700 cm$^{-1}$), the peaks of phonon transmission in borophene are much higher than those of graphene. These peaks are contributed by acoustic phonon modes. In addition, they are present for the transmission coefficients in both x and y directions, and the phonon transmission function along the x direction is similar to that along the y direction.

In the high frequency regime (700 cm$^{-1}$ ~ 1400cm$^{-1}$), there is a significant difference in the phonon transmission between the x and y direction: In the y direction, the transmission almost vanishes; while in the x direction, the transmission remains substantially high. Interestingly, the transmission coefficients in the x direction keep constant in the high frequency regime, except for a small central oscillation at about 1000cm$^{-1}$. Such feature of constant transmission resembles the phonon transmission along a perfect one-dimensional chain[55]. In principle, this phenomenon should not exist in 2D systems due to randomness of travelling directions of excited phonons and the interference of the phonon modes from different directions. Hence, this unusual behavior in borophene suggests that the travelling phonons in borophene are restricted solely in x direction.

This unusual phonon transmission is further evidenced by the mode-dependence analysis of the transmission coefficients. Figure 3(b) shows the x direction transmission coefficient and Figure 3(c) shows the y direction transmission coefficient, which are plotted against phonon frequencies and their respective transverse wave-vector. From Figure 3(b), it is obvious that in the high frequency regime, there exists a large area with the transmission coefficient $\Xi = 2$. This signifies that there exist two phonon modes travelling in the x direction, whose energy persists regardless of the transverse momentum. Such a feature disappears in the y direction transmission as shown in Figure3(c), where the phonon energy is proportional to phonon momentum. Hence the momentum of phonons is confined along the x direction. Due to the lack of contribution from high frequency phonon modes, the quantum thermal conductance in the y direction is lower than that along the x direction. As a result, the transport of high frequency phonon is oriented along the x direction, leading to the quantum thermal conductance along the x direction even higher than that of graphene.

**Atomistic origin of ultra-high thermal conductance in borophene**

In order to understand this unusual behavior of thermal conductance of borophene, we further investigate the wave function and atomistic coupling strength distribution of borophene using density functional theory (DFT) calculations. The electron density distribution from DFT calculation is shown in Figure 4(a). Interestingly, the isosurface plot shows that there is a strong anisotropic distribution in the electronic density: The electron density distribution is dense and one-dimensional along the B-B chain in the x direction, while there are obvious gaps between the chains along the y direction. In general, a denser electron distribution in-between the atoms should provide stronger interatomic force constants. From the density functional perturbation theory calculation, we are able to evaluate the dynamic matrix $D$ at each phonon mode $\boldsymbol{k}$. Then, the real space force-constant matrix $K$ can be obtained by the inverse Fourier transform of the dynamic matrix $D$ at $\boldsymbol{k}$ as following:

$$K_{ij}^{\alpha\beta}(\boldsymbol{R}) = \frac{1}{N_c}\sum_{\boldsymbol{k}} e^{i\boldsymbol{k}\cdot\boldsymbol{R}} D_{ij}^{\alpha\beta}(\boldsymbol{k}) \qquad (4)$$

where the $\alpha,\beta$ components correspond to the position vector of the $i$th, $j$th atoms in a unit cell, and $\boldsymbol{R}$ is the lattice vector in the real space and $N_c$ is the number of unit cells in the crystal. In NEGF calculation, we incorporate the IFC up to the 6$^{th}$ unit cell. For the purpose of illustration, we show in Figure 4(b) the strength of longitude interatomic force constant between nearest neighbor atoms in each direction. The strength of interatomic forces in the $x$ direction is about 30 times of that in the y direction, and 10 times of that between nearest buckling atoms. This implies that borophene has much stronger bonding in the x direction than in the y direction, making borophene like a collection of one-dimensional chains of boron atoms. Consequently, the cut-off frequency of phonon modes in the y direction, which is around 700cm$^{-1}$, is much smaller than that in the x direction. As a result, the phonons modes in the y direction (mostly in the low frequency regime) only interact with the low frequency modes in the x direction, and they nearly have no interactions with the high frequency modes in the x direction. Hence, the low frequency phonon transport exhibits 2D characteristics, while the high-frequency phonons transport shows one-dimensional characteristics. This unique bond strength arrangement leads to the extraordinarily high thermal conductance in the x direction of borophene, which is absent in other 2D materials.

**Thermal conductance of β12 borophene**

Recently, another atomic structure of single layer borophene, called β12 model, has been reported both experimentally[36] and theoretically[56]. The atomic structure of β12 model is shown in Figure 5(a). It is seen that no buckling is present in this model, similar to graphene. The lattice constants of structure, obtained from DFT calculation, are 5.062 Å in the x direction and 2.924Å in the y direction. Figure 5(b) shows the phonon dispersion relation of this structure experienced 1% tensile strain as suggested in Ref. 56. We find that the cut-off frequency is 1150cm$^{-1}$, which is even lower than that of graphene (1600 cm$^{-1}$). From the phonon density of states, we can observe that the low-frequency modes are much denser than high-frequency modes.

By using the same algorithm, we calculated the thermal conductance of β12 borophene, as shown in Figure 6. Since the thickness of β12 borophene has not been reported, we used the diameter of boron atom as the thickness, which is 1.81Å. We find that in β12 model, the thermal conductance is highly anisotropic, with the thermal conductance in *x* direction being much larger than that in *y* direction. The saturated thermal conductance in x direction is about 3 times of that in y direction. As shown in Figure 5(a), in x direction, such boron bond forms a boron chain globally. This boron chain greatly concentrates the phonons to travel in the same direction. On the other hand, such boron chain is absent in y direction. In comparison with graphene, a higher thermal conductance is realized in x direction.

In order to analyze the reason of high thermal conductance in β12 borophene, transmission function and electron density distribution are presented in Figure 7. We can find that the transmission function (Figure 7a) in the x direction is much greater than that in the y direction throughout. It is clear that the main contribution to the thermal conductance is from the low-frequency phonons (less than 600cm$^{-1}$). The first peak of transmission function near 200cm$^{-1}$ is mainly due to the acoustic modes, and the major peak at the 400cm$^{-1}$ is contributed both from acoustic and optical modes, but not the flexural modes. Hence this feature is quite different from graphene where the major contribution to thermal conductance is from the flexural modes. Figure 7(b) presents the electron density distribution. Unlike hexagonal borophene, the one-dimensional feature is not present, which is consistent with the absence of one-dimensional behavior in the phonon transmission function. Thus the mechanism of its high thermal conductance is different from that in hexagonal borophene. In β12 borophene, the presence of phonon transmission peak at 400cm$^{-1}$ is the major origin of its super-high thermal conductance, indicating that the low-frequency optical modes play a dominated role.

Table 1 summarizes the thermal conductance of borophene and graphene at room temperature. In order to eliminate the thickness effect, we also evaluated the per-length thermal conductance, where the thickness of 2D material is not divided. It is interesting to find that in x direction, the per-length thermal conductance of both hexagonal and β12 borophene sheets is higher than that of graphene, and the β12 borophene has a stronger anisotropy than its hexagonal counterpart.

## Conclusion

In the family of 2D materials, graphene was believed to have the highest thermal conductance and hence hold great promise for applications in heat dissipation and thermal management. Here, we report that both hexagonal and β12 borophene have comparable or even higher thermal conductance than graphene at room temperature. Except for the reasons of light atoms and strong interatomic bonds, this extraordinarily high thermal conductance is primarily rooted in the strong anisotropy of its lattice structure and the unusual phonon transmission. In hexagonal structured borophene, high-frequency phonons are only present in the x (ridge) direction but absent in the y (cross-ridge) direction. The mode selection along the x direction enables the thermal conductance of hexagonal borophene to be one-dimensional, thus suppressing randomness of phonon travelling directions in the high-frequency regime. In β12 borophene, the low frequency optical phonon modes contribute significantly to thermal conductance. These new mechanisms give rise to this extra-ordinarily high thermal conductance in monolayer borophene. The present study suggests that borophene is promising for applications in highly efficient heat dissipation and thermal management.

**ACKNOWLEDGMENT**

This work was supported in part by a grant from the Science and Engineering Research Council (152-70-00017). The authors gratefully acknowledge the financial support from the Agency for Science, Technology and Research (A*STAR), Singapore and the use of computing resources at the A*STAR Computational Resource Centre, Singapore.

## Author contributions

H. Z. performed NEGF calculations. Y. C. performed the first-principles calculations. G. Z. and Y.-W. Z. contributed to the discussion and writing of the manuscript. All authors reviewed the manuscript.

## Additional information

Competing financial interests: The authors declare no competing financial interests.

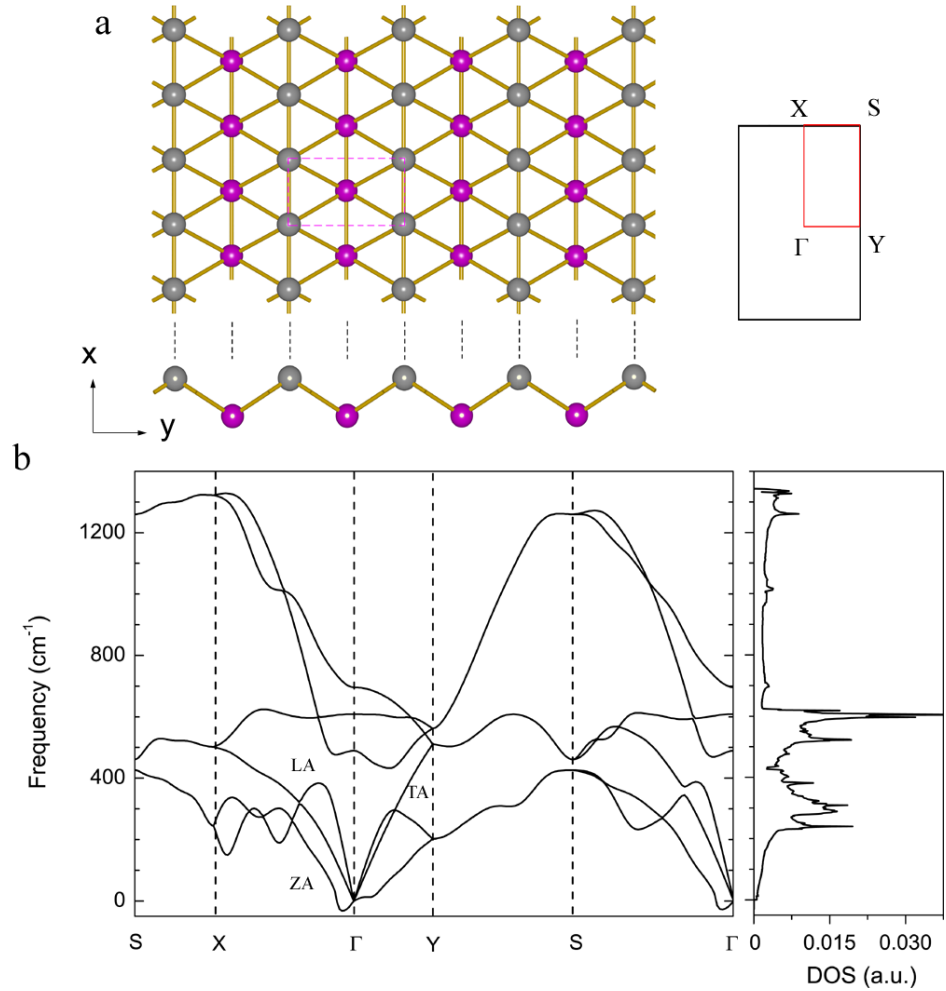

Figure 1. (a) The buckled crystal structure of borophene and the dashed lines show the unit cell. The atoms in upper and lower planes are represented by balls in grey and purple, respectively. The inset shows the Brillouin zone. (b) Phonon dispersion and density of states of borophene based on first-principles calculations.

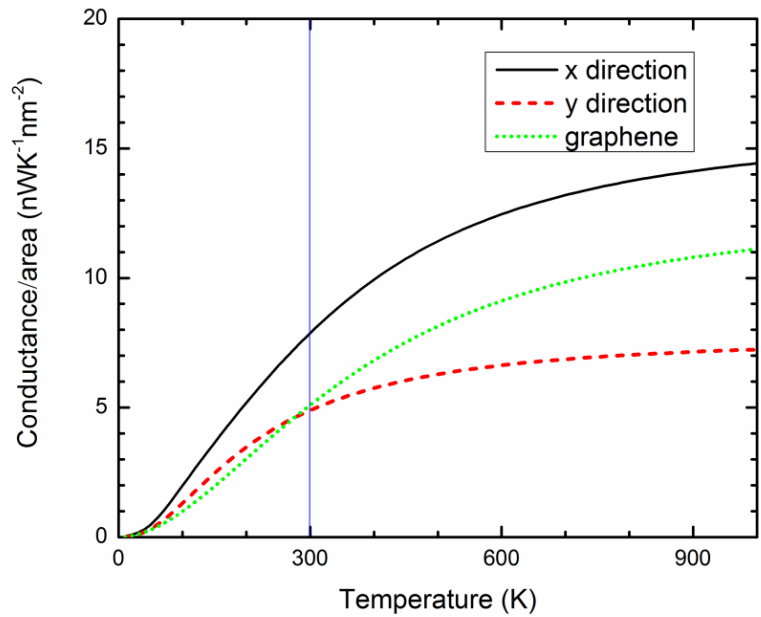

Figure 2.Variations of the thermal conductance per unit cross-sectional area of borophene with temperature in both x and y directions. The thermal conductance curve of graphene is also presented for comparison.

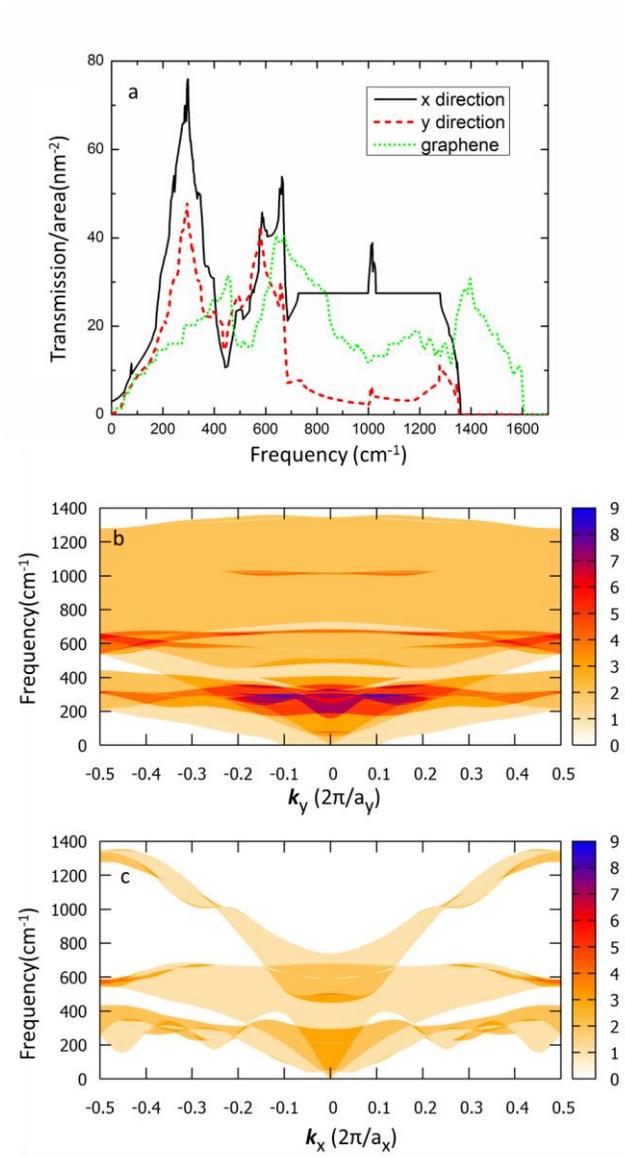

Figure 3. Transmission coefficient of phonons in borophene. (a) Plot of transmission coefficient $\mathcal{T}[\omega]$ per unit area against phonon frequency. (b) Plot of wavevector-dependent transmission coefficient $\Xi[\omega, k_y]$ of phonons travelling in the x direction against its transverse wave number $k_y$ and frequency $\omega$. (c) Plot of wave-vector dependent transmission coefficient $\Xi[\omega, k_x]$ of phonons travelling in the y direction against its transverse wave number $k_x$ and frequency $\omega$. Here $a_x$ and $a_y$ are the lattice constants in the x and y direction, respectively.

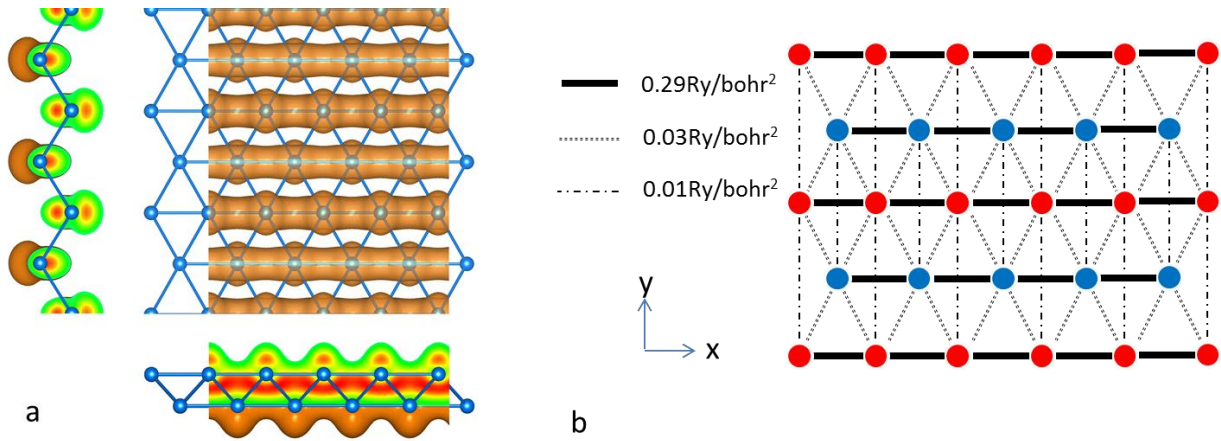

Figure 4. (a) Isosurface plot at $0.15\text{Å}^{-3}$ for electron density distribution of borophene in both top and side views. (b) Top view of interatomic force strength. The red atoms are in the upper plane and the blue ones are in the lower plane. The different types of lines illustrate the longitude interatomic force constants of the nearest neighbor atoms at each direction.

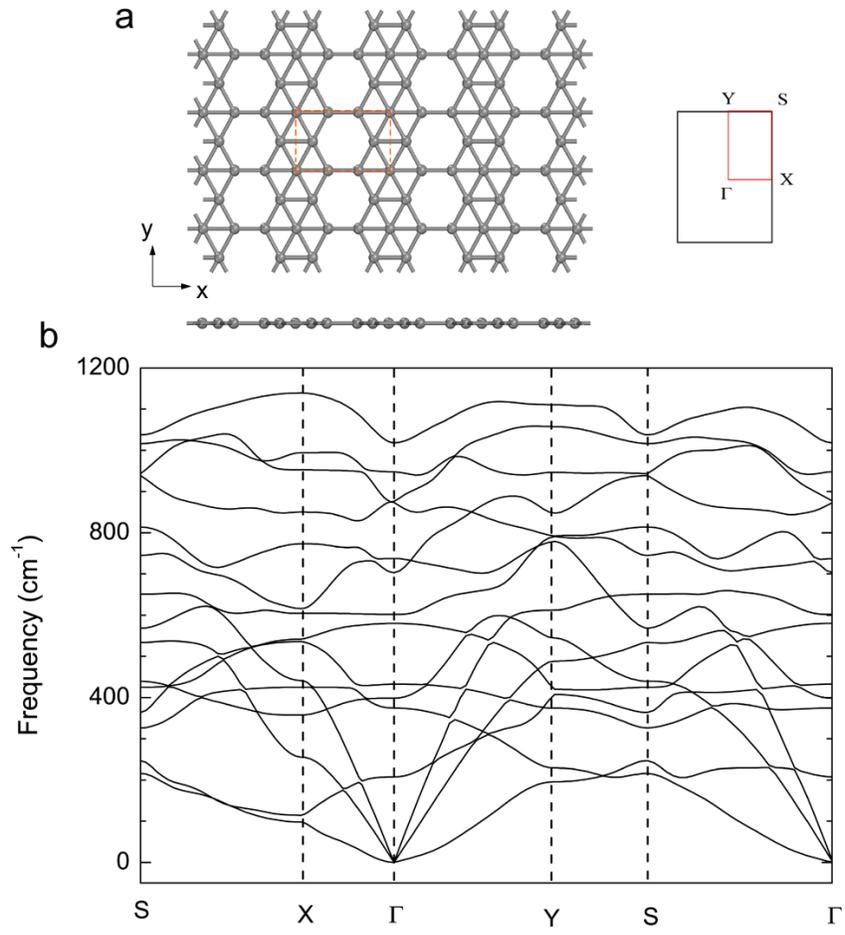

Figure 5. (a) Atomic model (top and side views) of the β12 borophene sheet. The unit cell is highlighted with dashed lines. The right panel shows the Brillouin zone. (b) Phonon dispersion of β12 borophene based on first-principles calculations.

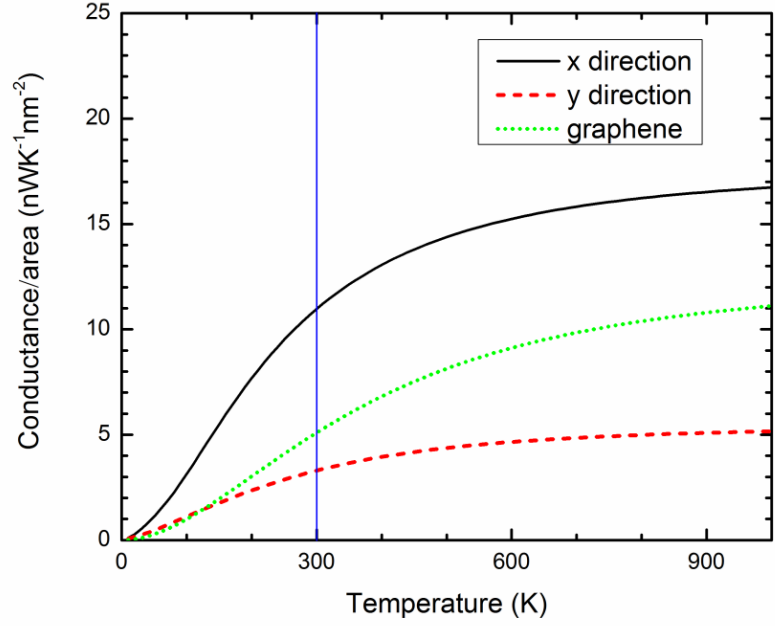

Figure 6. Temperature-dependent thermal conductance of β12 borophene and graphene. The thickness of borophene is taken as the diameter of boron atom. For graphene, the thickness is the interlayer distance (δ=3.4Å).

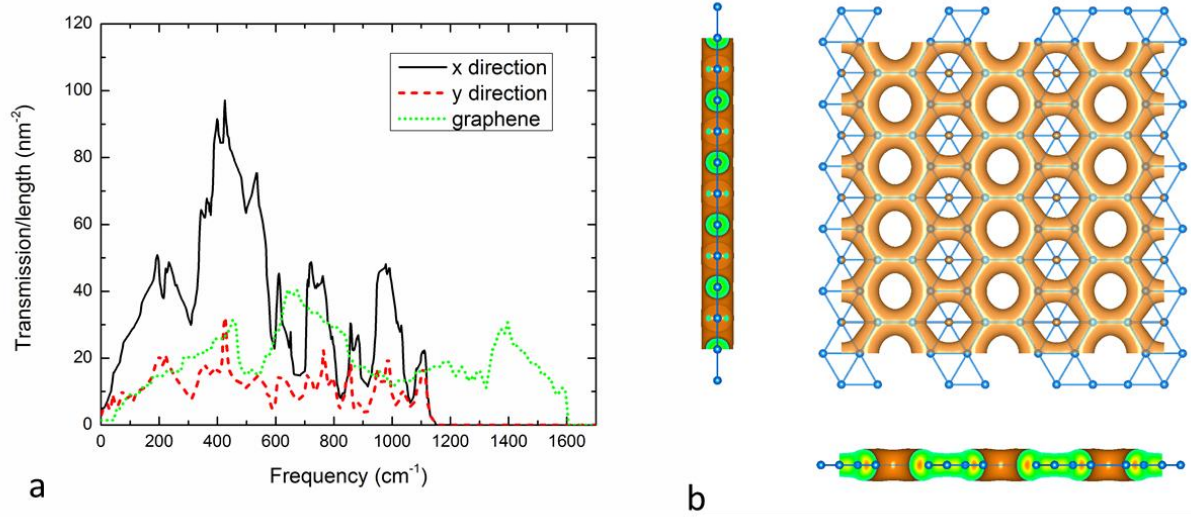

Figure 7. (a) Phonon transmission function of β12 borophene. (b) Isosurface plot at 0.15Å$^{-3}$ for electron density distribution of β12 borophene in both top and side views.

Table 1: A summary of thermal conductance of borophene and graphene at 300K. The per-area thermal conductance is defined as the heat current per temperature gradient per cross-sectional area, where the effect of thickness of 2D materials is counted. The thickness of borophene is taken as the atomic diameter plus the buckling height. Hence for the hexagonal model, δ=2.6Å and for the β12 model, δ=1.81Å. The per-length thermal conductance is defined as the heat current per temperature gradient per cross-sectional width, where the effect of thickness of 2D material is not counted.

|  |  | Per-area thermal conductance $(nW \cdot K^{-1} nm^{-2})$ | Per-length thermal conductance $(nW \cdot K^{-1} nm^{-1})$ |
|---|---|---|---|
| Borophene (hexagonal model) | x direction | 7.87 | 2.05 |
| | y direction | 4.89 | 1.27 |
| Borophene (β12 model) | x direction | 10.97 | 1.99 |
| | y direction | 3.30 | 0.60 |
| Graphene | isotropic | 5.10 | 1.71 |